\documentclass[aps,twocolumn,superscriptaddress,showpacs,notitlepage,float,nofootinbib]{revtex4-2}
\usepackage[utf8]{inputenc}
\usepackage{amsmath}
\usepackage{empheq}
\usepackage{amssymb}
\usepackage{mathbbol}
\usepackage{latexsym}
\usepackage{ifpdf}
\usepackage{slashed}
\usepackage{color}
\usepackage{mathtools}
\usepackage{soul}
\usepackage{hyperref}
\usepackage{mathrsfs}

\DeclareMathAlphabet{\mathpzc}{OT1}{pzc}{m}{it}

\ifpdf
\usepackage[pdftex]{graphicx}
\else
\usepackage{graphicx}
\fi

\ifpdf
\DeclareGraphicsExtensions{.pdf, .jpg, .tif}
\else
\DeclareGraphicsExtensions{.eps, .jpg}
\fi

\usepackage{tikz}

\def\XXint#1#2#3{{\setbox0=\hbox{$#1{#2#3}{\int}$ }
\vcenter{\hbox{$#2#3$ }}\kern-.6\wd0}}

\renewcommand{\underline}{\ul}
\newcommand{\bhat}[1]{\boldsymbol{\hat{#1}}}

\newlength\dlf  % Define a new measure, dlf

\begin{document}

\title{Calculation of bound and continuum states of the Ne$_{3}$ van der Waals trimer}

\author{Romain Guérout}
\email{romain.guerout@cnrs.fr}
\affiliation{Université Paris-Saclay, CNRS, Laboratoire Aimé Cotton, 91405, Orsay, France.}
% \date{\today}
\begin{abstract}
We use the configuration space Faddeev formalism to calculate bound and continuum states of the Ne$_{3}$ van der Waals trimer. Continuum states below the breakup threshold describe the scattering of a neon atom off of a Ne$_{2}$ diatomic molecule. We identify a resonant feature which we attribute to the presence of a three body resonance.
\end{abstract}

\maketitle

\section{Introduction} % (fold)
\label{sec:introduction}

One of the most popular theoretical methods for solving the quantum dynamics of few particles is the adiabatic hyperspherical formalism. We cannot cite the large body of work which has been done with this formalism. We refer the interested reader to a review article on the subject~\cite{greene2017universal}. In the context of atomic physics, a relatively less known formalism is the Faddeev method~\cite{faddeev1961scattering}. Originally developped in the context of nuclear physics, this formalism has been used both in configuration space~\cite{noyes1968calculations,payne1982configuration,merkuriev1976three,schellingerhout1989configuration} as well as in momentum space~\cite{mestrom2019finite,mestrom2020van}.

Recently, the configuration space Faddeev formalism has now been successfully applied to atomic physics problem. Of note is the use of this formalism for the calculation of a very loosely bound state of the H$_{2}^{+}$ molecular ion~\cite{carbonell2003new} without ever making use of the Born-Oppenheimer approximation. Most of the work done using the configuration space Faddeev formalism is concerned with the calculation of bound states of three particles although bound and scattering states of four particles are achievable numerically~\cite{lazauskas2006description}.

The quantum dynamics of three particles is a very rich subject which gives rise to a lot of different phenomena. The most notable one being of course the Efimov effect where an infinite series of loosely three body bound states appear at so-called ``universality windows'', when the scattering length associated to the pairwise interaction potential diverges. Outside of these universality windows, the Efimov bound states dissapear but three body resonances can subsist. Those correspond to three body quasi-bound states lying in the atom-dimer continuum. Those three body resonances have been studied for Coulombic~\cite{papp2005accumulation} or model~\cite{naidon2017efimov,happ2024mass} systems using methods based on complex scaling which allows to explore outside of the real energy axis.

In this work, we use the configuration space Faddeev method to perform atom-dimer scattering on the real energy axis for the neon van der Waals trimer. We first validate our method by reproducing previously calculated Ne$_{3}$ bound states. We identify a resonant feature in the atom-dimer scattering matrix which we attribute to the presence of a three body resonance. We then fully characterize this resonance by calculating the three body wavefunction. We note that the configuration space Faddeev formalism has been used previously for the study of the helium trimer~\cite{kolganova1998three}.

% section introduction (end)

\section{Theoretical formalism} % (fold)
\label{sec:theoretical_formalism}

We set $\hbar = 1$ and express energies $E$ as $E/k_{B}$ in kelvin (in the literature, energies are often expressed as $E/hc$ in cm$^{-1}$; we have 1 cm$^{-1}$ $\approx$ 1.43 K). We consider in the following a system of spinless particles. For a system of three particles of masses $m_1$, $m_2$ and $m_3$, we define the set of equivalent mass-scaled Jacobi vectors as
\begin{align}
    \mathbf{x}_{i} &= \sqrt{2 \mu_{jk}} \left(\mathbf{r}_{j} - \mathbf{r}_{k}\right) \\
    \mathbf{y}_{i} &= \sqrt{2 \mu_{i,jk}} \left(\mathbf{r}_{i} - \frac{m_{j} \mathbf{r}_{j} + m_{k} \mathbf{r}_{k}}{m_{j} + m_{k}}\right)
\end{align}
where, here and in the following, $(ijk)$ is a cyclic permutation of $(123)$, the $\mathbf{r}$ are the position vectors of the three particles and $\mu_{jk} = \frac{m_{j}m_{k}}{m_{j}+m_{k}}$, $\mu_{i,jk} = \frac{m_{i}(m_{j}+m_{k})}{m_{i}+m_{j}+m_{k}}$.
\begin{figure}[!h]
    \centering
    \includegraphics[width=0.4\textwidth]{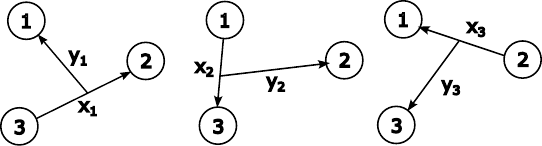}
    \caption{The three equivalent Jacobi coordinates for a system of three particles. Unscaled Jacobi vectors are shown.}
    \label{fig:equivJacobi}
\end{figure}
After separation of the motion of the total center of mass, the Schrödinger equation is 
\begin{equation}
    (\hat{T} + \hat{V} -E) \Psi(\mathbf{x}_{i},\mathbf{y}_{i}) = 0
\end{equation}
where the total wavefunction $\Psi$ can be written in any set of equivalent Jacobi coordinates. The Faddeev formalism~\cite{faddeev1961scattering} begins by expressing the total wavefunction $\Psi$ as a linear combination of Faddeev components $\phi_{i}$ each written in a given Jacobi coordinates
\begin{align}
    \label{eq:faddeev}
    \Psi(\mathbf{x}_{i},\mathbf{y}_{i}) =\phi_{1}(\mathbf{x}_{1},\mathbf{y}_{1}) + \phi_{2}(\mathbf{x}_{2},\mathbf{y}_{2}) + \phi_{3}(\mathbf{x}_{3},\mathbf{y}_{3})
\end{align}
% Upon the assumption that the three particles interact via pairwise potentials \emph{i.e.} $V = V_{1}(\mathbf{x}_{1}) + V_{2}(\mathbf{x}_{2}) + V_{3}(\mathbf{x}_{3})$ the Schrödinger equation for $\Psi$ becomes three coupled equations for the Faddeev components
The Schrödinger equation for $\Psi$ becomes three coupled equations for the Faddeev components
\begin{equation}
    \label{eq:faddeev}
    (\hat{T}_{i} + \hat{V}_{i} -E) \phi_{i}(\mathbf{x}_{i},\mathbf{y}_{i}) = -\hat{V}_{i}\left[ \phi_{j}(\mathbf{x}_{j},\mathbf{y}_{j}) + \phi_{k}(\mathbf{x}_{k},\mathbf{y}_{k}) \right]
\end{equation}
where $\hat{T}_{i}$ is the kinetic energy operator written in the Jacobi coordinates $(\mathbf{x}_{i},\mathbf{y}_{i})$
\begin{align}
    \hat{T}_{i} &= -\Delta_{\mathbf{x}_{i}}-\Delta_{\mathbf{y}_{i}} \nonumber \\
    &= -\frac{1}{x_{i}}\frac{\partial^{2}}{\partial x_{i}^{2}}x_{i} -\frac{1}{y_{i}}\frac{\partial^{2}}{\partial y_{i}^{2}}y_{i} + \frac{\hat{\ell}_{\mathbf{x}_{i}}^{2}}{x_{i}^{2}} + \frac{\hat{\ell}_{\mathbf{y}_{i}}^{2}}{y_{i}^{2}} \nonumber
\end{align}
where $\hat{\ell}_{\mathbf{x}_{i}}^{2}$ is the square angular momentum operator with respect to the angular part $\bhat{\mathbf{x}}_{i}$ of the Jacobi vector $\mathbf{x}_{i}$ and $\hat{V} = \sum_{i=1}^{3}\hat{V}_{i}$. Note that the differential equation for the component $\phi_{i}$ only involves the potential $V_{i}$ and that eq.~\eqref{eq:faddeev} is really three equations through cyclic permutations of $(ijk)$.

In order to numerically solve the coupled differential equations for the Faddeev components, we follow the procedure described in~\cite{schellingerhout1989configuration}. The Faddeev components are first expanded on a basis of bipolar spherical harmonics $\mathcal{Y}_{\alpha}(\bhat{\mathbf{x}}_{i},\bhat{\mathbf{y}}_{i})$
\begin{align}
    \label{eq:partial}
    \phi_{i}(\mathbf{x}_{i},\mathbf{y}_{i}) &= \sum_{\alpha}^{} \frac{f_{i,\alpha}(x_{i},y_{i})}{x_{i} y_{i}} \mathcal{Y}_{\alpha}(\bhat{\mathbf{x}}_{i},\bhat{\mathbf{y}}_{i}) \\
    \mathcal{Y}_{\alpha}(\bhat{\mathbf{x}}_{i},\bhat{\mathbf{y}}_{i}) &= \left[  Y_{\ell_{\alpha}}(\bhat{\mathbf{x}}_{i}) \otimes Y_{\lambda_{\alpha}}(\bhat{\mathbf{y}}_{i}) \right]_{LM} \\
    \hat{\ell}_{\mathbf{x}_{i}}^{2} \mathcal{Y}_{\alpha}(\bhat{\mathbf{x}}_{i},\bhat{\mathbf{y}}_{i}) &= \ell_{\alpha}(\ell_{\alpha}+1) \mathcal{Y}_{\alpha}(\bhat{\mathbf{x}}_{i},\bhat{\mathbf{y}}_{i}) \\
    \hat{\ell}_{\mathbf{y}_{i}}^{2} \mathcal{Y}_{\alpha}(\bhat{\mathbf{x}}_{i},\bhat{\mathbf{y}}_{i}) &= \lambda_{\alpha}(\lambda_{\alpha}+1) \mathcal{Y}_{\alpha}(\bhat{\mathbf{x}}_{i},\bhat{\mathbf{y}}_{i})    
\end{align}
where $\alpha$ collects angular momentum quantum numbers $(\ell_{\alpha},\lambda_{\alpha})$ which couple to give a particular value of the total angular momentum $L$ and will be called in the following a \emph{channel}.
Taking advantage of the orthogonality relation of the bipolar spherical harmonics, we obtain an infinite number of coupled integro-differential equations for the partial Faddeev components $f_{i,\alpha}$
\begin{widetext}
    \begin{equation}
    \label{eq:partialFaddeev}
        (t_{i,\alpha}-E)f_{i,\alpha}(x_{i},y_{i})+\sum_{\beta}^{}v_{i,\alpha\beta} \, f_{i,\beta}(x_{i},y_{i}) = -\sum_{\beta,\gamma}^{} v_{i,\alpha\gamma} \iint d \bhat{\mathbf{x}}_{i}d \bhat{\mathbf{y}}_{i} \frac{x_{i}y_{i}}{x_{j}y_{j}} \mathcal{Y}_{\gamma}^{*}(\bhat{\mathbf{x}}_{i},\bhat{\mathbf{y}}_{i}) f_{j,\beta}(x_{j},y_{j}) \mathcal{Y}_{\beta}(\bhat{\mathbf{x}}_{j},\bhat{\mathbf{y}}_{j})
    \end{equation}
\end{widetext}
where $t_{i,\alpha} = -\frac{\partial^{2}}{\partial x_{i}^{2}} - \frac{\partial^{2}}{\partial y_{i}^{2}} + \frac{\ell_{\alpha}(\ell_{\alpha}+1)}{x_{i}^{2}} + \frac{\lambda_{\alpha}(\lambda_{\alpha}+1)}{y_{i}^{2}}$ and $v_{i,\alpha\beta}$ is the channel interaction term
\begin{equation}
    v_{i,\alpha\beta}(x_{i},y_{i}) = \iint d \bhat{\mathbf{x}}_{i}d \bhat{\mathbf{y}}_{i} \, \mathcal{Y}_{\alpha}^{*}(\bhat{\mathbf{x}}_{i},\bhat{\mathbf{y}}_{i}) 
   V_{i}(\mathbf{x}_{i},\mathbf{y}_{i}) \mathcal{Y}_{\beta}(\bhat{\mathbf{x}}_{i},\bhat{\mathbf{y}}_{i})
\end{equation}
We have written here the most general form of the channel interaction potential when the functions $V_{i}$ depends on both $\mathbf{x}_{i}$ and $\mathbf{y}_{i}$. The fact that $V_{i}$ depends on $\mathbf{y}_{i}$ means that the interaction between particle $j$ and $k$ depend on the position of the third particle $i$. This can occur when taking into account non-additive three body interaction. The fact that $V_{i}$ depends on $\mathbf{x}_{i}$ can occur when the interaction between particle $j$ and $k$ is anisotropic \emph{e.g.} in the case of a dipolar interaction.

In the following, we do not take into account any non-additive three body term and consider an isotropic interaction between neon atoms. As a consequence, the $V_{i}$ only depends on $x_{i} = \left\| \mathbf{x}_{i} \right\|$ and can readily be identified as pairwise interactions.

Finally, the right-hand side of eq.~\eqref{eq:partialFaddeev} is the so-called Jacobi kernel which couples the different Faddeev components. Note that the Jacobi kernel is a purely geometrical term. 

In order to numerically solve eq.~\eqref{eq:partialFaddeev}, the partial Faddeev components are further expanded onto a basis of cubic Hermite spline functions $s_{n}$ as
\begin{equation}
    f_{i,\alpha}(x_{i},y_{i}) = \sum_{n,m}^{} a^{i,\alpha}_{nm}s_{n}(x_{i})s_{m}(y_{i})
\end{equation}
Such an expansion in conjonction with an orthogonal collocation method leads to the construction of the different operators appearing in eq.~\eqref{eq:partialFaddeev} : the kinetic energy operator $\mathbb{T}$, the channel interaction operator $\mathbb{V}$, an indicator operator $\mathbb{1}$ and the Jacobi kernel $\mathbb{K}$. Explicit expressions for the matrix elements of those operators are given in the appendix.

For bound states calculation, after imposing that the wavefunction must vanish at infinity, the problem is reduced to the generalized eigenvalue problem $\left(\mathbb{T} + \mathbb{V} + \mathbb{K} \right)v = E \mathbb{1} v$ where $v$ are the solution vectors collecting the expansion coefficients $a^{i,\alpha}_{nm}$ and $E$ are the bound states energies.

For a scattering calculation~\footnote{We consider only scattering below the three body breakup.}, we use the formalism of the forced Schrödinger equation and the unknown partial Faddeev component $f_{i,\alpha}$ is split into an incoming and a scattered part i.e. $f_{i,\alpha} = f^{inc}_{i,\alpha} + f^{sca}_{i,\alpha}$. The incoming part is a known asymptotic solution for $\mathbb{K} = 0$. The resolution of the eqs.~\eqref{eq:partialFaddeev} reduces then to the resolution of the linear system of equations $(\mathbb{T} + \mathbb{V} + \mathbb{K} - E \mathbb{1}) v = -\mathbb{K} \chi$  where $\chi$ is a vector of expansion coefficients of the incoming part $f^{inc}_{i,\alpha}$. In a multichannel seting, the corresponding equation is 
\begin{equation}
    \mathbf{f}_{i} = \mathbf{f}_{i}^{inc} + \mathbf{f}_{i}^{sca}
\end{equation}
where $\mathbf{f}_{i}$ collects $N$ linearly independent solutions of the scattering problem with $N$ the number of open asymptotic scattering channels at a given total energy $E$. 

As a base pair of asymptotic channels function, we take the functions $\{\varphi_{v,j}^{i}(x_{i}) \, j^{i}_{\varepsilon}(y_{i}) , \varphi_{v,j}^{i}(x_{i}) \, h^{i}_{\varepsilon}(y_{i})\}$ where $\varphi_{v,j}^{i}(x_{i})$ is the wavefuction of a rovibrational state of the dimer with energy $E_{v,j}$. Although we denote this state with a rotational quantum number $j$ as customary it should be clear that we have in fact $j = \ell_{\alpha}$ given the previous discussion. The functions $j^{i}_{\varepsilon}(y_{i})$ and $h^{i}_{\varepsilon}(y_{i})$ are regular Ricatti-Bessel and outgoing Ricatti-Hankel functions~\footnote{We have $h^{i}_{\varepsilon} = -n^{i}_{\varepsilon} + i j^{i}_{\varepsilon}$ in terms of regular and irregular Ricatti-Bessel functions $\{j^{i}_{\varepsilon},n^{i}_{\varepsilon}\}$.} respectively at energy $\varepsilon$. The energies are related by $E = E_{v,j} + \varepsilon$. In a multichannel setting, we collect all the regular functions in a vector $\mathbf{j}$ and all the outgoing functions in a vector $\mathbf{h}$. We therefore set $\mathbf{f}^{inc}_{i} = \mathbf{1} \, \mathbf{j}$ in order to get a set of $N$ linearly independent solutions. Asymptotically \emph{i.e.} as $y_{i} \to \infty$, the Jacobi kernel vanishes~\cite{merkuriev1976three} so that the scattered partial Faddeev components behave as a linear combination of the asymptotic channels functions
\begin{equation}
     \mathbf{f}_{i}^{sca} \xrightarrow[y_{i} \to \infty]{} \mathbf{X} \, \mathbf{j} + \mathbf{Y} \, \mathbf{h}
 \end{equation}
 with $\mathbf{X}$ and $\mathbf{Y}$ the matrices obtained by projecting each scattered partial Faddeev component onto each asymptotic channels functions. Analyzing the solutions in terms of regular and outgoing asymptotic channels functions in this way leads to the transition matrix $\mathbf{T}$ from which the scattering matrix $\mathbf{S}$ can be obtained
 \begin{align}
     \mathbf{T} &= (\mathbf{1} + \mathbf{X})^{-1} \, \mathbf{Y} \\
     \mathbf{S} &= \mathbf{1} + 2 i \mathbf{T}
 \end{align}

% section theoretical_formalism (end)

\section{Application} % (fold)
\label{sec:application}

We apply this formalism to the characterization of the bound and continuum states of Ne$_{3}$. As a consequence of the indiscernability of the three neon atoms, the Faddeev formalism is modified in the sense that there is a unique Faddeev component $\phi$. Nevertheless, the total wavefunction is still given by 
\begin{equation}
    \Psi(\mathbf{x}_{i},\mathbf{y}_{i}) =\phi(\mathbf{x}_{1},\mathbf{y}_{1}) + \phi(\mathbf{x}_{2},\mathbf{y}_{2}) + \phi(\mathbf{x}_{3},\mathbf{y}_{3})
\end{equation}
and most of the general formalism presented above still apply.

For the interaction potential $V(x)$ between two neon atoms, we take the analytical form given in~\cite{wuest2003determination} 
\begin{align}
    V(x) &= A e^{-b x} - \sum_{n=3}^{8} f_{2n}\left(x,b\right) \frac{C_{2n}}{x^{2n}} \\
    f_{2n}\left(x,b\right) &= 1 - e^{-b x} \sum_{k=0}^{2n} \frac{\left(b x\right)^{k}}{k!}
\end{align}
which consists of a repulsive barrier together with an attractive dispersion tail. This analytical form for the Ne-Ne potential was fitted to reproduce measured rovibrational states of Ne$_{2}$. 

We use $m(^{20}\text{Ne}) = 19.992\,440\,1753$ Da~\cite{wang2021ame} in our calculations. The grid in $x$ consists in 50 points ranging from 0 to 50 a$_{0}$. The grid is non uniform to allow for more points at short range and is chosen so as to reproduce correctly the rovibrational states of the dimer Ne$_{2}$. The same grid has been used for both $x$ and $y$. Calculations are done for the $L^{\pi} = 0^{+}$ symmetry. The angular basis consists of $\ell_{\alpha} = \lambda_{\alpha}$ with $\ell_{\alpha}$ even and ranging from 0 to 10.

\subsection{Bound states of Ne$_{3}$} % (fold)
\label{sub:subsection_name}

In this section, we present our results for the calculation of the bound states energies of the trimer Ne$_{3}$. We recall that we have to solve the generalized eigenvalue problem $\mathbb{A} v = E \mathbb{1} v$ where $\mathbb{A} = \mathbb{T} + \mathbb{V} + \mathbb{K}$ consisting of the kinetic energy operator, the channel interaction potential and the Jacobi kernel. In doing so, it is important to never have to evaluate $\mathbb{1}^{-1}$ in order to take advantage of the sparsity and the banded structure of the indicator operator $\mathbb{1}$ which collects the values of the cubic Hermite spline basis functions at the collocation points. We employ the following inverse iteration method: given initial guesses $E_{0}$ and $v_{0}$ for the bound state energy and vector of expansion coefficients, we define the recursion relation
\begin{equation}
    \label{eq:recursion}
    (E_{0} \mathbb{1} - \mathbb{A}) v_{k} = \mathbb{1} v_{k-1}
\end{equation}
Then, the sequence $\frac{\langle v_{k} | v_{k-1} \rangle}{\langle v_{k} | v_{k} \rangle}$ converges towards $E_{0} - E_{i}$ where $E_{i}$ is a generalized eignevalue \emph{i.e.} a bound state energy. Note that the recursion relation~\eqref{eq:recursion} is a system of linear equations for the unkown $v_{k}$ as the operator $E_{0} \mathbb{1} - \mathbb{A}$ as well as the solution vector $\mathbb{1} v_{k-1}$ are given. This method is largely insensitive to the initial guess $v_{0}$ for the expansion coefficient vector which we then take as a random vector.

We show in Table~\ref{tab:Ne3BS} our results for the bound state energies of Ne$_{3}$. We compare our results to those of Ref.~\cite{suno2019study} which uses an adiabatic hyperspherical treatment. The calculations from Ref.~\cite{suno2019study} use an interaction potential whose parameters are taken from Ref.~\cite{tang2003van} which slightly differ from the parameters from Ref.~\cite{wuest2003determination} which we use. 

To try to quantify those differences, in Ref.~\cite{suno2019study} the ground state energy of Ne$_{2}$ as well as the scattering length are respectively $-24.1305$ K and $32.3$ a$_{0}$. With the potential we use, we obtain $-24.2428$ K and $28.99$ a$_{0}$ for those same quantities.

\begin{table}[h]
    \caption{Bound states energies of Ne$_{3}$ for $L^{\pi} = 0^{+}$. Energies are in units of $K$ and relative to the three body breakup threshold. The present results are compared with those based of an adiabatic hyperspherical treatment (Ref.~\cite{suno2019study}) which uses a slightly different potential (see text for details).\\ }
    \label{tab:Ne3BS}
    \centering

    \begin{tabular}{lllll}
    \hline
    \hline

    & & This work & & Ref.~\cite{suno2019study} \\
    \hline
        $n=0$ & \; & $-73.12$ & \; & $-73.17$ \\
        $1  $ &    & $-51.62$ &    & $-51.63$ \\
        $2  $ &    & $-48.66$ &    & $-48.55$ \\
        $3  $ &    & $-44.70$ &    & $-44.77$ \\
        $4  $ &    & $-39.38$ &    & $-39.55$ \\
        $5  $ &    & $-33.57$ &    & $-33.82$ \\
        $6  $ &    & $-31.79$ &    & $-31.62$ \\
        $7  $ &    & $-30.70$ &    & $-30.82$ \\
        $8  $ &    & $-26.78$ &    & $-26.98$ \\
        $9  $ &    & $-24.94$ &    & $-25.42$ \\
        $10 $ &    & $-24.54$ &    & $-24.60$ \\

    \hline
    \hline
    \end{tabular}
\end{table}

\begin{figure}[h]
    \centering
    \includegraphics[width=0.5\textwidth]{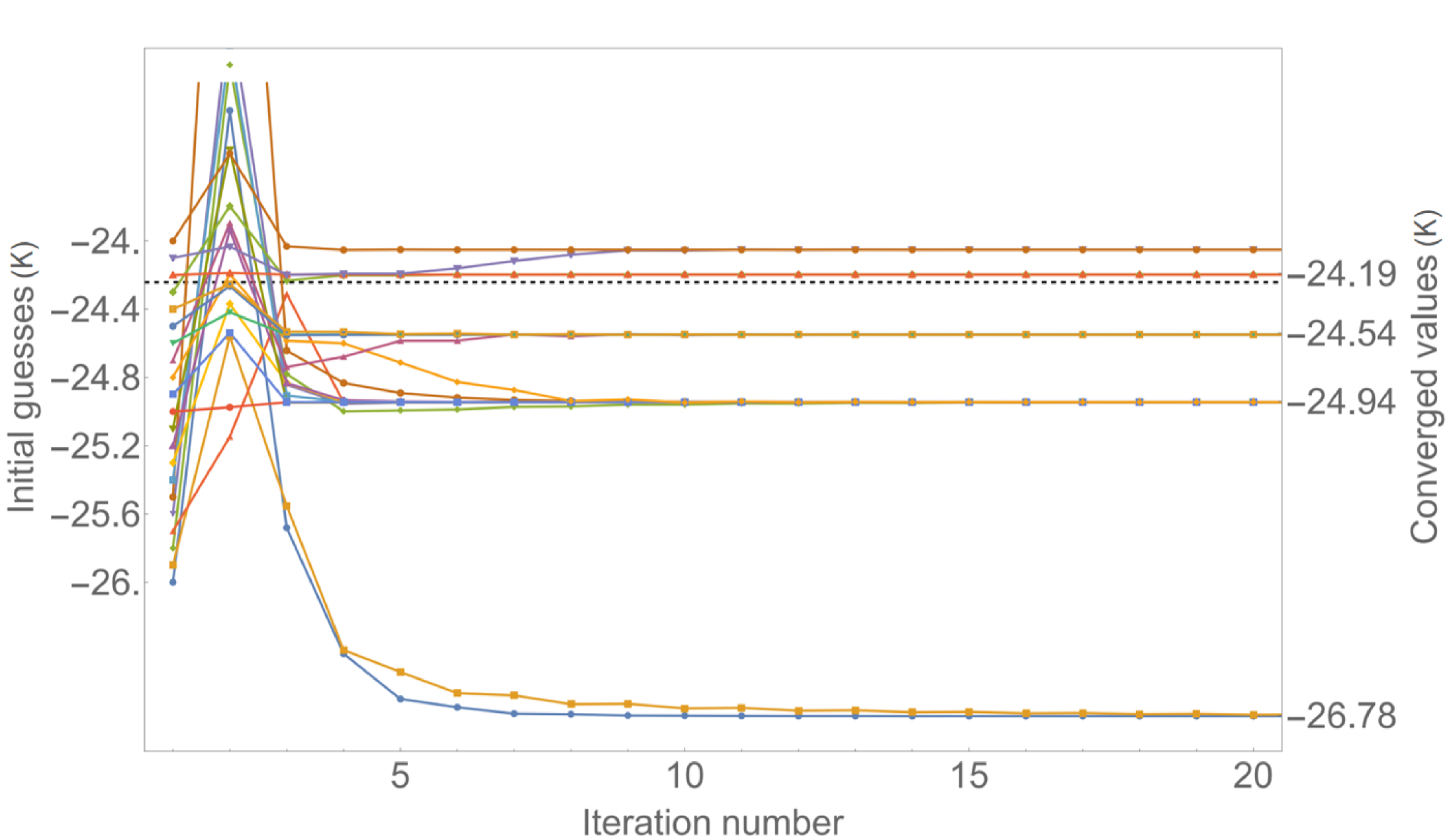}
    \caption{Convergence of the few least bounded states of Ne$_{3}$ via the inverse iteration method. Starting from a uniform set of initial guesses between $-26$ K and $-24$ K (left side of the graph), converged bound states energies as a function of the iteration number (right side of the graph). The ground state of Ne$_{2}$ is shown as a black dashed horizontal line.}
    \label{fig:ne3bsconv}
\end{figure}

We show in Fig.~\eqref{fig:ne3bsconv} the convergence of the few least bounded states of Ne$_{3}$ to illustrate our inverse iteration method. We can see how, starting from a uniform range of initial guesses, the method indeed converges towards several bound state energies in a few iterations. In this energy range, we can also see convergence towards states above the ground state energy of Ne$_{2}$: those states are the first few so-called ``box states'' which result from the discretization of the continuum as a consequence of the boundary conditions being imposed on a finite ``box'' $(x,y) \in [0,50 \text{ a}_{0}] \otimes [0,50 \text{ a}_{0}]$.

% subsection subsection_name (end)

\subsection{Continuum states of Ne$_{3}$} % (fold)
\label{sub:continuum_states_of_ne}

At energies $E$ greater than the ground state energy of Ne$_{2}(v=0,j=0)$, we are in the atom-dimer continuum. In this situation, the resolution of the Schrödinger equation after applying the collocation method involves solving the linear system of equation $\mathbb{A}v = -\mathbb{K} \chi$ with $\mathbb{A} = \mathbb{T} + \mathbb{V} + \mathbb{K} -E \mathbb{1}$. Furthermore, in the context of the forced Schrödinger equation formalism, this linear system is inhomogeneous with a source term of the form $-\mathbb{K} \chi$ where $\chi$ is a vector of expansion coefficients for a particular solution with $\mathbb{K} = 0$. This particular solution is easy to obtain: when setting $\mathbb{K} = 0$, the Schrödinger equation separates in an equation in $x$ and an equation in $y$. A particular solution has the form $\varphi_{v,j}(x) j_{\varepsilon}(q y)$ consisting of a bound state of Ne$_{2}$ with energy $E_{v,j}$ and a Ricatti-Bessel function for the free Ne atom. Conservation of energy gives
\begin{equation}
    E = E_{v,j} + \varepsilon = E_{v,j} + \frac{q^{2}}{2 \mu_{1,23}}
\end{equation}
with $\mu_{1,23} = \frac23 m$ being the reduced mass in the $y$ direction. Let $b$ be a vector collecting the values of $\varphi_{v,j}(x) j_{\varepsilon}(q y)$ at the collocation points, the source term $\chi$ is then the solution of the linear system $\mathbb{1} \chi = b$.

Analysis of the scattered solutions in terms of asymptotic forms consisting of Ricatti-Bessel and outgoing Ricatti-Hankel functions gives the transition matrix $\mathbf{T}$ and the scattering matrix $\mathbf{S}$. Important properties of the scattering matrix are unitarity (which expresses the conservation of probabilities) and reciprocity (which expresses time reversal symmetry). Those are never imposed in our formalism so checking those properties constitutes a stringent numerical test. We define $\eta_{U}$ and $\eta_{R}$ the defects from unitarity and reciprocity respectively as
\begin{align}
    \eta_{U} &= \left\| \mathbf{1} - \mathbf{S}\mathbf{S}^{\dagger} \right\| \\
    \eta_{R} &= \frac{\left\| \mathbf{S} - \mathbf{S}^{\mathsf{T}} \right\|}{\left\| \mathbf{S} + \mathbf{S}^{\mathsf{T}} \right\|}
\end{align}
where $\|.\|$ denote any matrix norm which we take as the 2-norm.

\begin{figure}[!h]
    \centering
    \includegraphics[width=0.5\textwidth]{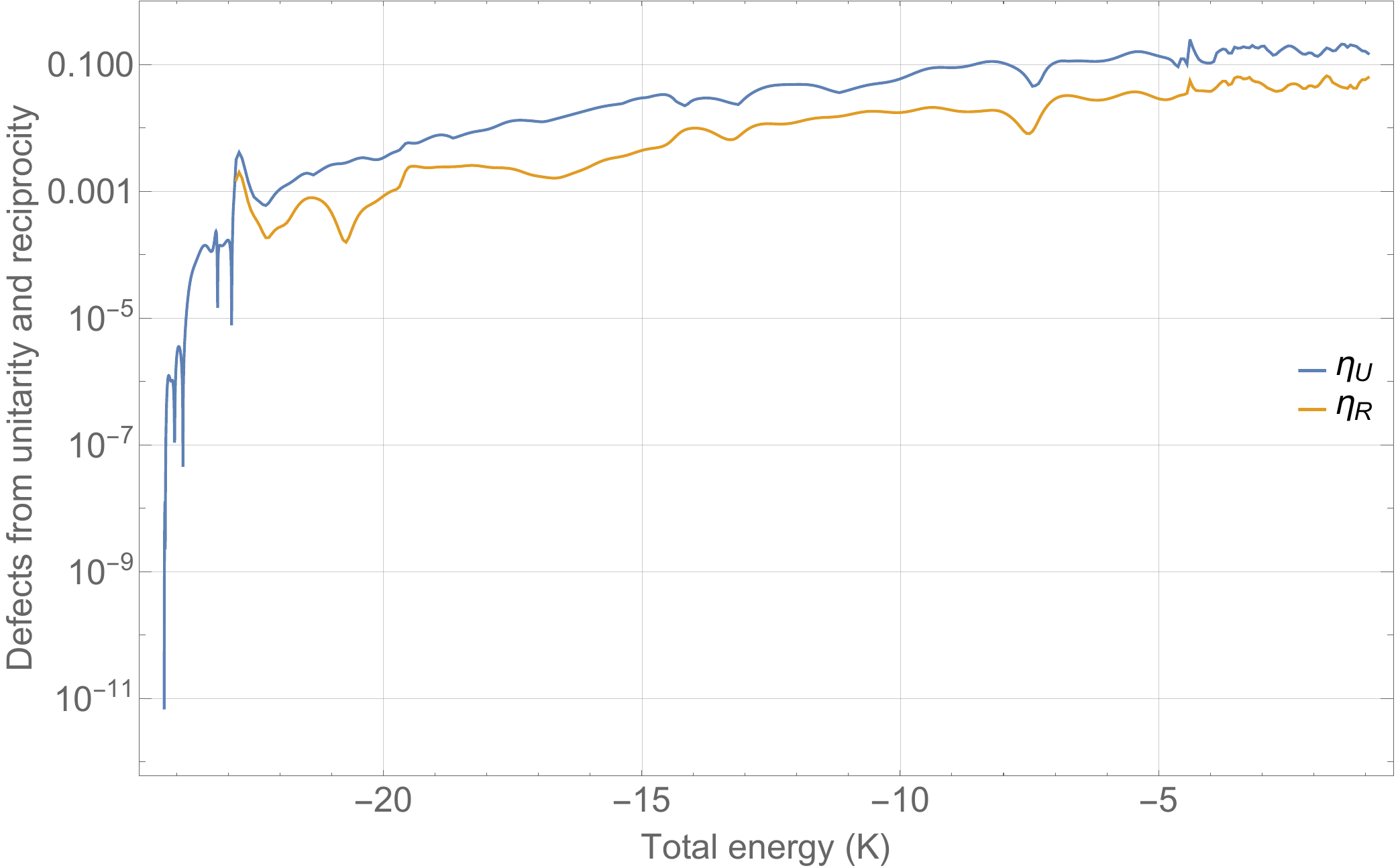}
    \caption{Defects from unitarity and reciprocity of the calculated scattering matrix as a function of the total energy.}
    \label{fig:defects}
\end{figure}

We show in Fig.~\eqref{fig:defects} the calculated quantities $\eta_{U}$ and $\eta_{R}$ as a function of the total energy from the onset of the atom-dimer continuum up to the three body breakup threshold.

For a given number of grid points and basis functions in the $x$ and $y$ directions, a maximum energy can be reliably described. That is why we see in Fig.~\eqref{fig:defects} that the values of both $\eta_{U}$ and $\eta_{R}$ steadily increase as the total energy increases. At the opening of the first inelastic channel, $\eta_{U}$ is of the order of $10^{-3}$ which is very satisfactory. Near the three body breakup threshold, both quantities are around $1-10$ $\%$ and we have reached the limits of the current grid.

\begin{figure}[!h]
    \centering
    \includegraphics[width=0.5\textwidth]{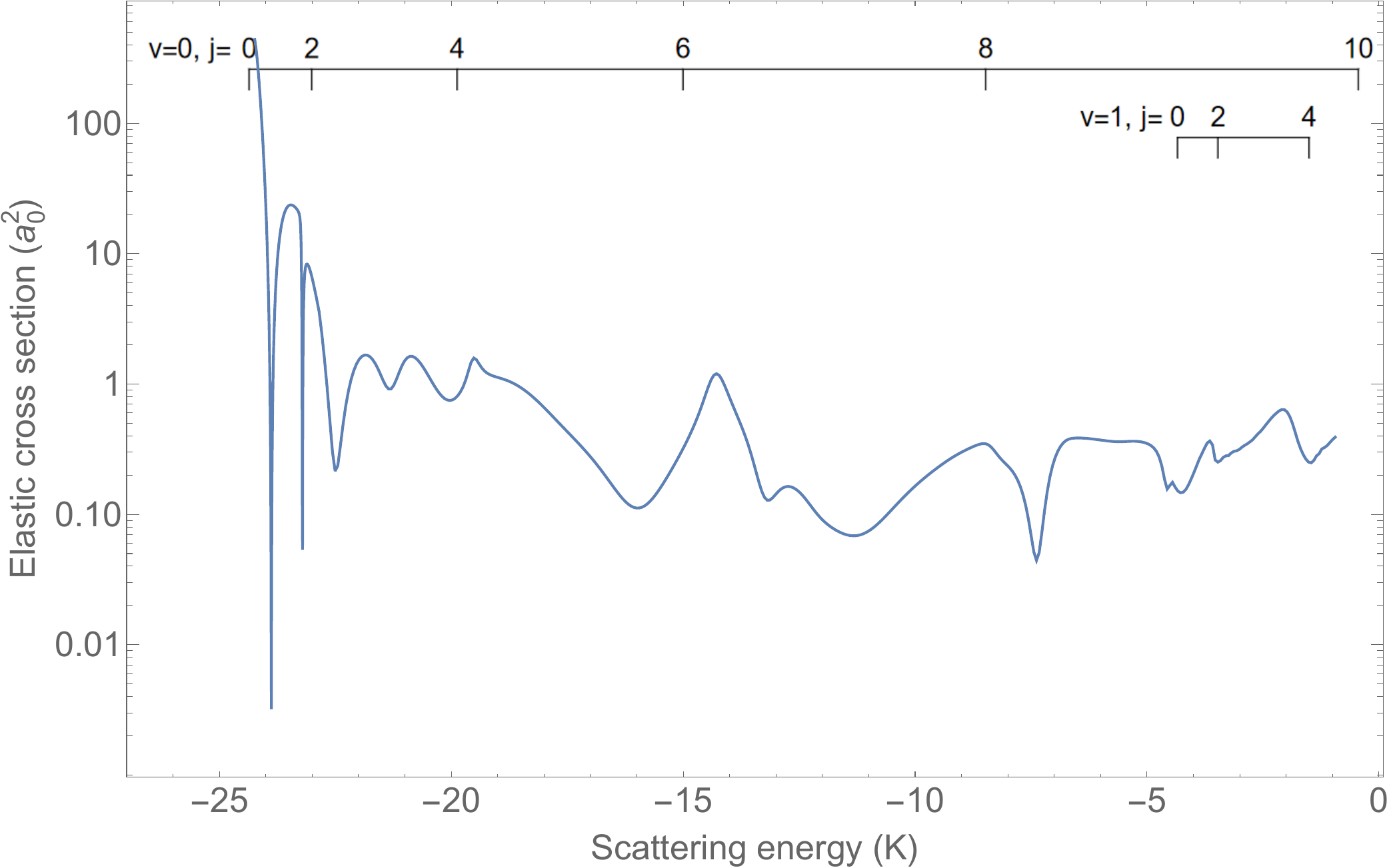}
    \caption{Elastic cross section for a Ne atom off of a Ne$_{2}(v=0,j=0)$ molecule.}
    \label{fig:Ne3Scattering}
\end{figure}

We show in Fig.~\eqref{fig:Ne3Scattering} the elastic cross section for a Ne atom off of a Ne$_{2}$ molecule in its ground state $v=0$, $j=0$.
If we index the $\mathbf{T}$ matrix elements according to the state of the Ne$_{2}$ molecule i.e. $\mathbf{T}_{vj,v'j'}$, the elastic cross section $\sigma_{el}$ shown in Fig.~\eqref{fig:Ne3Scattering} is
\begin{equation}
    \sigma_{el} = \frac{\pi}{q^{2}} | \mathbf{T}_{00,00} |^{2}
\end{equation}

\begin{figure}[!h]
    \centering
    \includegraphics[width=0.5\textwidth]{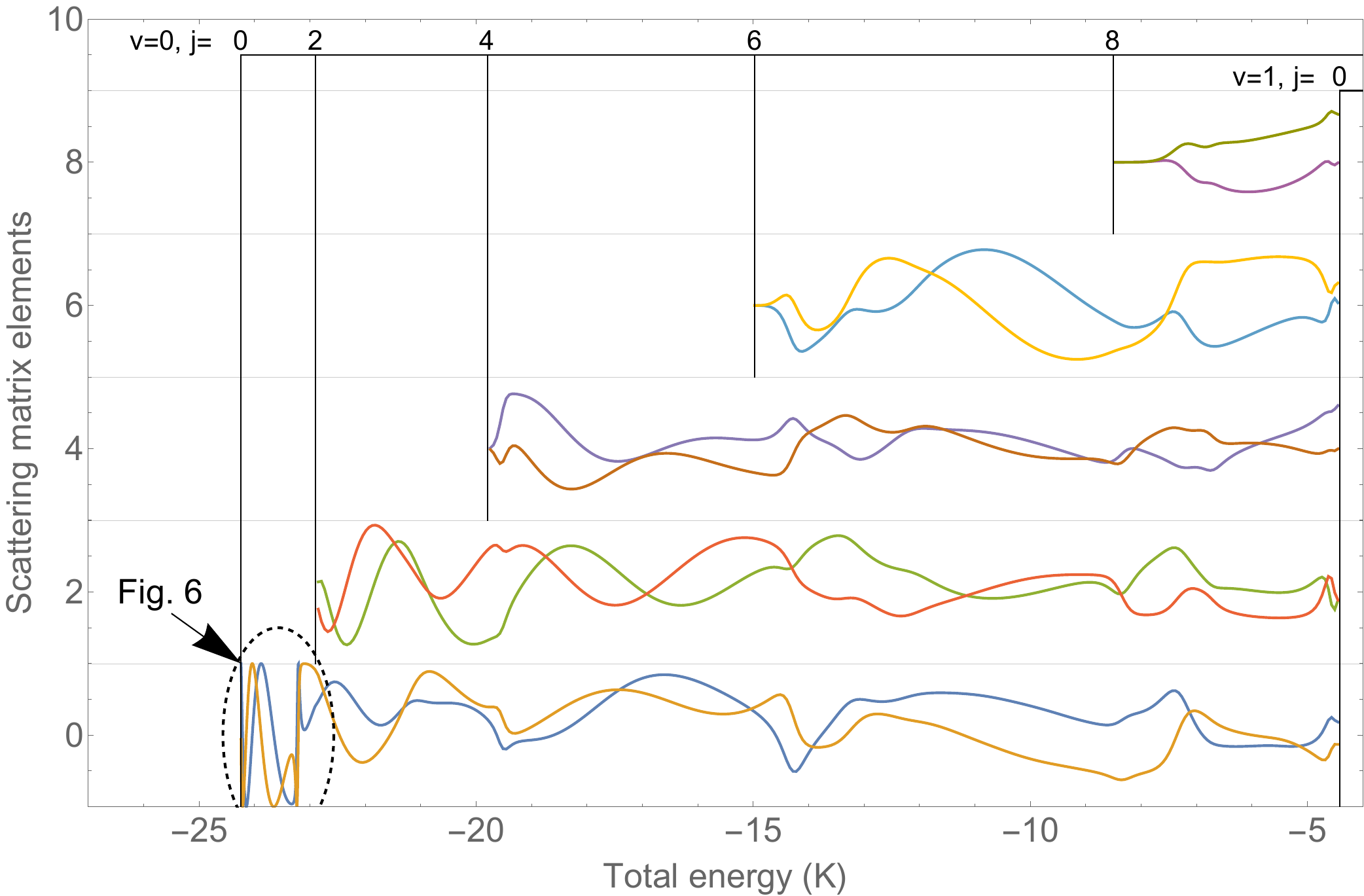}
    \caption{Scattering matrix elements for a Ne atom off of a Ne$_{2}(v=0,j=0)$ molecule. The matrix elements have been shifted horizontally for clarity. Each matrix element can be identified according to where it opens up in energy.}
    \label{fig:Ne3ScatteringElements}
\end{figure}

Furthermore, we show in Fig.~\eqref{fig:Ne3ScatteringElements} the various scattering matrix elements up to the first vibrational excitation threshold. Each scattering matrix element has been shifted horizontally for clarity.

 The figure shows the real and imaginary part of each matrix element. The lowest blue and orange curves correspond to the elastic scattering matrix element $\mathbf{S}_{11}$ for the process $\text{Ne}_{2}(v=0,j=0) + \text{Ne}(q) \to \text{Ne}_{2}(v=0,j=0) + \text{Ne}(q)$, the red and green curves above them correspond to the first inelastic scattering process $\text{Ne}_{2}(v=0,j=0) + \text{Ne}(q) \to \text{Ne}_{2}(v=0,j=2) + \text{Ne}(q')$ correspnding to the $\mathbf{S}_{12}$ matrix element and so on. Note that threshold laws are readily visible in Fig.~\eqref{fig:Ne3ScatteringElements} where inelastic scattering elements for processes $\text{Ne}_{2}(v=0,j=0) + \text{Ne}(q) \to \text{Ne}_{2}(v=0,j') + \text{Ne}(q')$ behave as $q'^{j'+1}$~\cite{gullans2017efimov}.

% subsection continuum_states_of_ne (end)

\subsection{Three body resonance} % (fold)
\label{sub:three_body_resonances}

\begin{figure}[!h]
    \centering
    \includegraphics[width=0.5\textwidth]{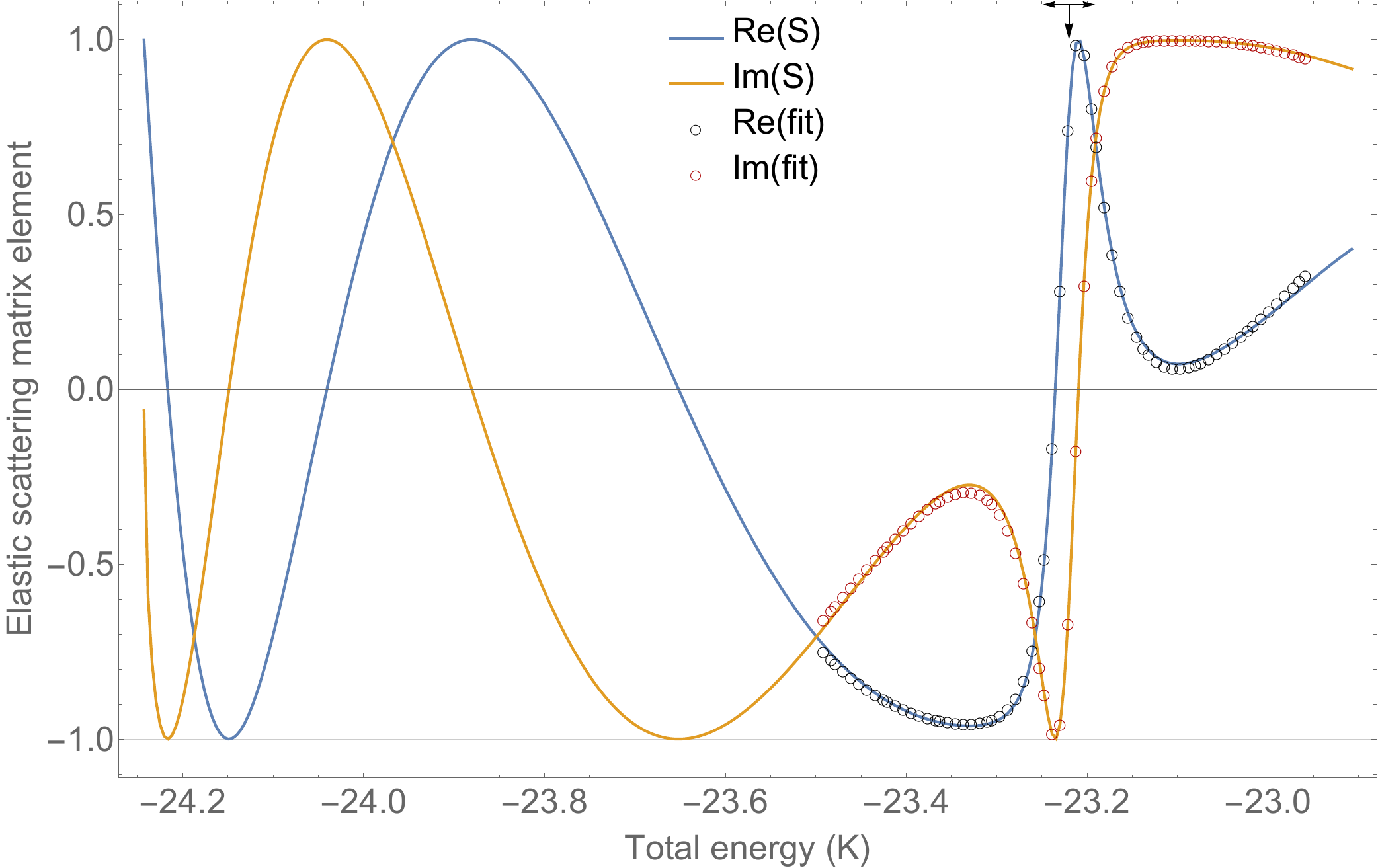}
    \caption{Elastic scattering matrix element for a Ne atom off of a Ne$_{2}(v=0,j=0)$ molecule. This corresponds to the energy region highlighted in Figure~\ref{fig:Ne3ScatteringElements}. Also shown is the best fit with a resonant form (open circles, see text for details). Position and width of the fitted resonance is indicated by arrows.}
    \label{fig:Ne3ScatteringElementsResonance}
\end{figure}

We show in Fig.~\eqref{fig:Ne3ScatteringElementsResonance} the elastic scattering matrix element in the energy region below the first inelastic threshold. This corresponds to the energy region highlighted in Fig.~\eqref{fig:Ne3ScatteringElements}.

We can clearly identify a resonant feature which we attribute to the presence of a three body resonance. In order to calculate the energy of this resonance, $E_{r}$, we fit the scattering matrix element with the following model in a small energy window around the resonance
\begin{align}
    \mathbf{S}(E) &\approx S_{bg}(q) \frac{E-E_{r}^{*}}{E-E_{r}} \\
    S_{bg}(q) &= e^{i(\alpha q + \beta)}
\end{align}
This model consists of a non resonant background $S_{bg}(q)$ which we take as a unit modulus function of an affine phase and a unitary resonant part~\cite{racsseev1992resonant} describing the presence of a pole of the scattering matrix in the complex plane. The best fit of the elastic scattering matrix element with this model is shown in Fig.~\eqref{fig:Ne3ScatteringElementsResonance} and leads to a three body resonance energy
\begin{equation}
    E_{r} = (-23.22-i 0.029) \text{K}
\end{equation}
The width of this resonance $\Gamma = 2 |\text{Im}(E_{r})|$ leads to a lifetime $\tau \approx 0.13 $ ns. We note that in this energy region, the scattering matrix is a single channel quantity which probably explain why this three body resonance is so well resolved. 

\begin{figure*}
    \centering
    \includegraphics[width=\textwidth]{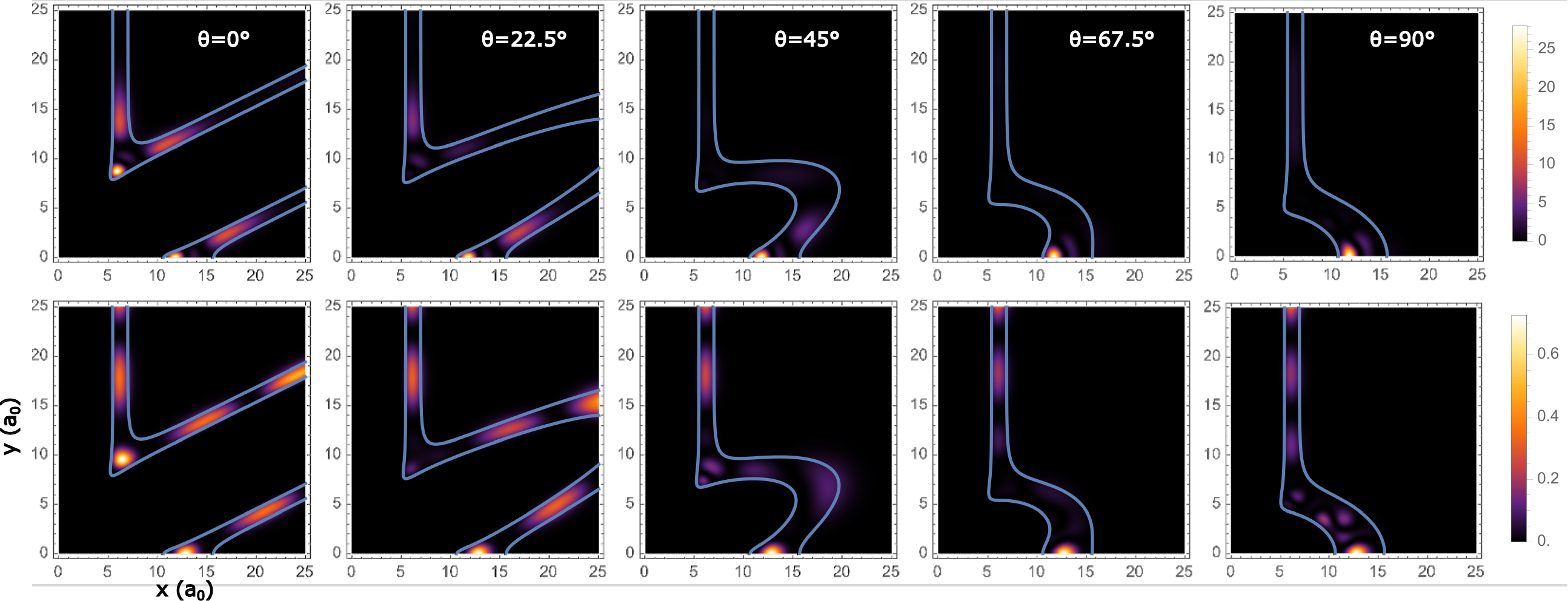}
    \caption{Total continuum wavefunction $\Psi$ of Ne$_{3}$ at the energy of the three body resonance (top row) and off resonance (bottom row), in Jacobi coordinates, for selected Jacobi angles $\theta$. The quantity represented is $|\rho^{2}\Psi|^{2}$ with $\rho$ the hyperradius. Unscaled Jacobi vectors corresponding to true distances in a$_{0}$ are used. The blue contour delimits the classically allowed region $V=E$. Note the difference in scale in the color code between the top and bottom rows.}
    \label{fig:ResWF}
\end{figure*}

We show in Fig.~\eqref{fig:ResWF} the total continuum wavefunction $\Psi$ of Ne$_{3}$ at the energy of the three body resonance and slightly off resonance. Given the expression of the expansion of the Faddeev components in terms of partial channels (see eq.~\eqref{eq:partial}) and the fact that the hyperradius $\rho$ is independent of the Jacobi coordinates used, a factor $\rho^{-2}$ can be factored out of $\Psi$. Therefore, we represent in Fig.~\eqref{fig:ResWF} the quantity $|\rho^{2} \Psi|^{2}$ to enhance the features of the wavefunction at large $\rho$. The top row of Fig.~\eqref{fig:ResWF} represent the wavefunction at the energy of the resonance (to be precise at $E=-23.208$ K chosen such that the scattered part of the wavefunction vanishes) while the bottom row represents the wavefunction slightly off resonance (at $E=-23.5$ K) for comparison. The first thing to note is the difference in the scale of the color code between the top row and the bottom row. Both wavefunctions (on and off resonance) are normalized in the same way in terms of their asymptotic form $\varphi_{0,0}(x)\left[ j_{\varepsilon}(y) + T_{11} h_{\varepsilon}(y) \right]$ so comparing the two makes sense. The vast difference in the scale of the color code confirms that the wavefunction at resonance exhibits a large probability density at short range where a transient Ne$_{3}$ state is formed.

The wavefunction is shown in Jacobi coordinates $(x_{1},y_{1})$ at selected Jacobi angle $\theta$. At each Jacobi angle, the blue contour delimits the classically allowed region $V=E$. Examining the bottom row, we see that the off resonance wavefunction is as expected: it explores the whole classically allowed region without any significant peaks in the probability density. On the contrary, the wavefunction at resonance exhibits very high density of probability at certain geometries. There is a strong probability density at the linear configuration corresponding to $y_{1}=0$. As it can be better seen at $\theta=0^{\circ}$, this linear configuration is rather floppy with a non negligeable probability density for $y_{1} \neq 0$ which is also present at other Jacobi angles~\footnote{Note that particle indiscernability is readily visible for $\theta=0^{\circ}$.}. We can then conclude that the geometry of this three body resonance of Ne$_{3}$ correspond to a somewhat linear configuration with a Ne atom in a broad region between the other two Ne atoms. We note however that for $\theta=90^{\circ}$ we do not see any probability density at the equilateral triangle geometry. This resonance state is not floppy enough to explore that configuration.

\begin{figure}[tb]
    \centering
    \includegraphics[width=0.5\textwidth]{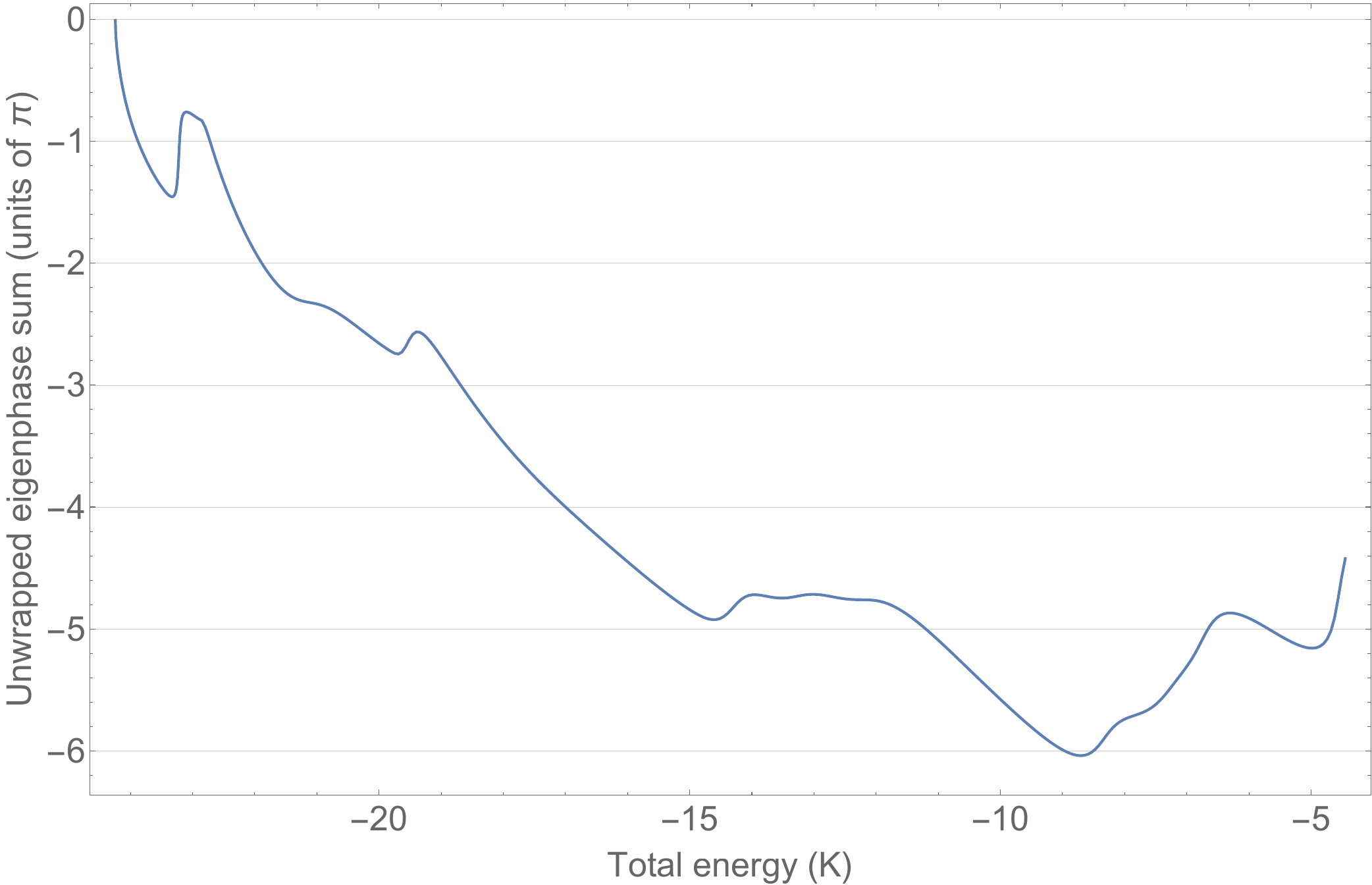}
    \caption{Unwrapped eigenphase sum (in units of $\pi$) up to the first vibrational excitation threshold. This corresponds to the same energy range as in Fig.~\eqref{fig:Ne3ScatteringElements}.}
    \label{fig:eigenphase}
\end{figure}

As the energy increases and the scattering matrix becomes a multichannel quantity, the identification of other resonances usually rely on the eigenphase sum $\Delta = \sum_{n} \delta_{n}$ where the $e^{2 i \delta_{n}}$ are the eigenvalues of the scattering matrix. It is well known~\cite{hazi1979behavior} that the eigenphase sum $\Delta$ shows a characteristic jump of $\pi$ around an isolated resonance. We show in Fig.~\eqref{fig:eigenphase} the eigenphase sum up to the first vibrational excitation threshold.
Unfortunately, it is difficult to unambiguously identify other three body resonances at higher energies. Studying Figs.~\eqref{fig:Ne3ScatteringElements} and~\eqref{fig:eigenphase} we can only tentatively report on broad features around $E \approx -19.5$ K, $E \approx -14.5$ K and $E \approx -7.5$ K. None of these features show a clear jump of $\pi$ in the eigenphase sum though so we cannot say further.

% subsection three_body_resonances (end)

% section application (end)

\section{Conclusion} % (fold)
\label{sec:conclusion}

Using a configuration space Faddeev formalism, we have calculated bound and continuum states of the Ne$_{3}$ van der Waals trimer. We have identify and fully characterize a three body resonance. Our work confirms that standard scattering on the real energy axis can be an alternative to the use of complex scaling methods to describe three body resonances.

% section conclusion (end)

\section*{Acknowledgements} % (fold)
\label{sec:acknowledgements}

I dedicate this work to my doctoral advisor Dr. Martin Jungen (1938–2024, University of Basel, Switzerland).

% section acknowledgements (end)

\section*{Appendix: Orthogonal collocation method} % (fold)
\label{sec:appendix_orthogonal_collocation_method}

Here, we consider unscaled Jacobi coordinates; hence reduced masses appear explicitly in the kinetic energy operator matrix elements. After defining a maximum value $x_{max}$ in the $x$ direction, subdivide the interval $[0,x_{max}]$ into $n_{x}/2$ parts. This defines $n_{x}$ collocation points and $n_{x}/2+1$ grid points. Let $\mathpzc{x}_{i}$ and $x_{i}$ denote those collocation and grid points respectively. Each subinterval $[x_{i},x_{i+1}]$ contains 2 collocation points defined as the nodes of a 2-points Gaussian quadrature on this subinterval.

Cubic Hermite splines functions $s_{n}(x)$ come in two types and are relative to a given grid point $x_{n}$. They satisfy the following properties
\begin{align}
    \text{Type I: }  & s^{\text{I}}_{n}(x_{n})=1, d s^{\text{I}}_{n}/dx|_{x=x_{n}}=0 \\
    \text{Type II: } & s^{\text{II}}_{n}(x_{n})=0, d s^{\text{II}}_{n}/dx|_{x=x_{n}}=1
\end{align}
and the support of $s_{n}(x)$ is $[x_{n-1},x_{n+1}]$. These properties are used to impose a specific logarithmic derivative at a certain grid point; indeed let $f(x) = s^{\text{I}}_{n}(x) + b s^{\text{II}}_{n}(x)$. Then $\frac{d \ln f}{d x}|_{x=x_{n}} = b$.

All cubic spline functions can be defined from primitive ones, defined on $[0,2]$ relative to a parameter $0 \leq a \leq 2$
\begin{align}
    \text{Type I: } & s^{\text{I}}(x)= \begin{cases}
        \frac{(3a-2x)x^{2}}{a^{3}} & 0 \leq x < a \\
        \frac{(x-2)^2 (3a-2(x+1))}{(a-2)^3} & a \leq x \leq 2
    \end{cases} \\
    \text{Type II: } & s^{\text{II}}(x)= \begin{cases}
        (x-a)\frac{x^2}{a^2} & 0 \leq x < a \\
        (x-a)\frac{(x-2)^{2}}{(a-2)^{2}} & a \leq x \leq 2
    \end{cases}
\end{align}
through appropriate scaling and translation.

The same apply in the $y$ direction over an interval $[0,y_{max}]$ with $n_{y}$ colloction points. Then, the two dimensional basis function $s_{n}(x)s_{m}(y)$ is defined relative to a grid point $(x_{n},y_{m})$ and its support is the rectangle whose four corners are $(x_{n \pm 1},y_{m \pm 1})$ which we call a grid tile.

The kinetic energy operator $\mathbb{T}$, the indicator operator $\mathbb{1}$ and the channel interaction operator $\mathbb{V}$ are all local in the sense that for a given collocation point $(n,m)$, in a given Jacobi coordinates $i$, $(\mathpzc{x}_{i,n},\mathpzc{y}_{i,m})$ it requires the value of said operator at this collocation point. 

A given collocation point belongs at most to 4 different grid tiles and at most 4 different two dimensional basis functions will have this grid tile as a support. It follows that all local operators will have at most 16 non-zero matrix element per lines leading to a sparse and banded structure. In addition, those operators benefit from a factorized structure with respect to the Kronecker product $\otimes$. For instance, we have for the indicator operator $\mathbb{1}$
\begin{align}
    \left[ \mathbf{s}_{x} \right]_{nm} &= s_{m}(\mathpzc{x_{n}}) \\
    \left[ \mathbf{s}_{y} \right]_{nm} &= s_{m}(\mathpzc{y_{n}}) \\
    \mathbb{1} &= \mathbf{1}_{n_{c}} \otimes \mathbf{1}_{n_{\alpha}} \otimes \mathbf{s}_{y} \otimes \mathbf{s}_{x}
\end{align}
where $n_{c}$ and $n_{\alpha}$ are the number of Faddeev components and the number of channels. While spectral methods have an overlap operator between basis functions, pseudo-spectral methods have an operator collecting the values of the basis functions at the grid points. In our collocation method, the $\mathbb{1}$ operator collects the values of the basis functions at the collocation points. Furthermore, since we use basis functions having a compact support, this operator can be viewed as indicating whether a given collocation point belongs to the support of a given basis function. Hence the analogy with the indicator function.

For the channel interaction operator $\mathbb{V}$ in the case of isotropic interaction, we have
\begin{align}
    \left[ \mathbf{v}_{i} \right]_{nm} &= V_{i}(\mathpzc{x_{n}}) s_{m}(\mathpzc{x_{n}}) \\
    \mathbb{v}_{i} &= \mathbf{1}_{n_{\alpha}} \otimes \mathbf{s}_{y} \otimes \mathbf{v}_{i} \\
    \mathbb{V} &= \begin{pmatrix}
        \mathbb{v}_{1} & \mathbb{0} & \mathbb{0} \\
        \mathbb{0} & \mathbb{v}_{2} & \mathbb{0} \\
        \mathbb{0} & \mathbb{0} & \mathbb{v}_{3}
    \end{pmatrix}
\end{align}

For the kinetic energy operator, 
\begin{align}
    \left[ \mathbf{s}''_{x} \right]_{nm} &= s''_{m}(\mathpzc{x_{n}}) \\
    \left[ \mathbf{s}''_{y} \right]_{nm} &= s''_{m}(\mathpzc{y_{n}}) \\
    \mathbf{d_{x^{2}}} &= \mathbf{1}_{n_{\alpha}} \otimes \mathbf{s}_{y} \otimes \mathbf{s}''_{x}\\
    \mathbf{d_{y^{2}}} &= \mathbf{1}_{n_{\alpha}} \otimes \mathbf{s}''_{y} \otimes \mathbf{s}_{x}\\
    \left[\boldsymbol{\ell}_{\alpha}\right]_{nm} &= \frac{\ell_{\alpha}(\ell_{\alpha}+1)}{\mathpzc{x}_{n}^2} s_{m}(\mathpzc{x}_{n}) \\
    \left[\boldsymbol{\lambda}_{\alpha}\right]_{nm} &= \frac{\lambda_{\alpha}(\lambda_{\alpha}+1)}{\mathpzc{y}_{n}^2} s_{m}(\mathpzc{y}_{n}) \\
    \boldsymbol{\ell}_{x} &= \begin{pmatrix}
        \mathbf{s}_{y} \otimes \boldsymbol{\ell}_{1} & & & \\
         & \mathbf{s}_{y} \otimes \boldsymbol{\ell}_{2} & & \\
         & & \ddots & \\
         & & & \mathbf{s}_{y} \otimes \boldsymbol{\ell}_{n_{\alpha}}
    \end{pmatrix} \\
    \boldsymbol{\ell}_{y} &= \begin{pmatrix}
        \boldsymbol{\lambda}_{1} \otimes \mathbf{s}_{x} & & & \\
         & \boldsymbol{\lambda}_{2} \otimes \mathbf{s}_{x} & & \\
         & & \ddots & \\
         & & & \boldsymbol{\lambda}_{n_{\alpha}} \otimes \mathbf{s}_{x}
    \end{pmatrix} \\
    \mathbb{t}_{i} &= -\frac{1}{2 \mu_{jk}} \left(\mathbf{d_{x^{2}}} - \boldsymbol{\ell}_{x} \right) -\frac{1}{2 \mu_{i,jk}} \left(\mathbf{d_{y^{2}}} - \boldsymbol{\ell}_{y} \right) \\
    \mathbb{T} &= \begin{pmatrix}
    \mathbb{t}_{1} & \mathbb{0} & \mathbb{0} \\
    \mathbb{0} & \mathbb{t}_{2} & \mathbb{0} \\
    \mathbb{0} & \mathbb{0} & \mathbb{t}_{3}
    \end{pmatrix}
\end{align}

At first glance, the Jacobi kernel operator $\mathbb{K}$ has no sparsity nor structure due to its non-local nature: for a given collocation point $(n,m)$, in a given Jacobi coordinates $i$, $(\mathpzc{x}_{i,n},\mathpzc{y}_{i,m})$ it requires the evaluation of the kernel at points $(x_{j},y_{j})$, in other Jacobi coordinates $j$. However, the value of the hyperradius $\rho^{2} = \mu_{jk} x_{i}^{2} + \mu_{i,jk} y_{i}^2$ is independent of the Jacobi coordinates used. As a consequence, for a given collocation point $(\mathpzc{x}_{i,n},\mathpzc{y}_{i,m})$, the values taken by the points $(x_{j},y_{j})$ lie on a elliptical arc corresponding to the values of the hyperradius. This arc will intercept only a finite number of grid tiles leading to some degree of sparsity.

A basic ingredient in evaluating the Jacobi kernel is the expression of the Jacobi coordinates $(\mathbf{x}_{j},\mathbf{y}_{j})$ and $(\mathbf{x}_{k},\mathbf{y}_{k})$ as a function of $(\mathbf{x}_{i},\mathbf{y}_{i})$. This can be done simply by inspection of Fig.~\eqref{fig:equivJacobi} and we get
\begin{subequations}
\label{eqs:jacobis}
\begin{align}
    -\mathbf{x}_{j} &= \frac{m_{j}}{m_{j}+m_{k}} \mathbf{x}_{i} + \mathbf{y}_{i} \\
     \mathbf{y}_{j} &= \frac{m_{i}}{m_{i}+m_{k}} \mathbf{x}_{j} + \mathbf{x}_{i} \\
    -\mathbf{x}_{k} &= \frac{m_{k}}{m_{j}+m_{k}} \mathbf{x}_{i} - \mathbf{y}_{i} \\
    -\mathbf{y}_{k} &= \frac{m_{i}}{m_{i}+m_{j}} \mathbf{x}_{k} + \mathbf{x}_{i}
\end{align}
\end{subequations}

In absence of external fields, the problem has rotational invariance. As such, the Jacobi vector $\mathbf{x}_{i}$ can be chosen to lie along the $z$ axis of a laboratory reference frame and we have
\begin{equation}
    \iint d \bhat{\mathbf{x}}_{i}d \bhat{\mathbf{y}}_{i} \to 8 \pi^{2} \int_{0}^{\pi} d \theta \sin \theta 
\end{equation}
where $\theta = \widehat{(\mathbf{x}_{i},\mathbf{y}_{i})}$ is the polar angle of the Jacobi vector $\mathbf{y}_{i}$. We then have for the Jacobi kernel matrix element for Faddeev components $(i,j)$, channels $(\alpha,\beta)$, collocation points $(nm,pq)$

\begin{widetext}
\begin{equation}
    \left[\mathbb{K}\right]_{i\alpha nm,j \beta pq} = (1-\delta_{ij}) 8 \pi^{2} V_{i}(\mathpzc{x}_{i,m}) \int_{0}^{\pi} d \theta \sin \theta \frac{\mathpzc{x}_{i,m} \mathpzc{y}_{i,n}}{x_{j} y_{j}} \mathcal{Y}_{\alpha}^{*}(0,0,\theta,0) s_{q}(x_{j}) s_{p}(y_{j}) \mathcal{Y}_{\beta}(\theta_{\mathbf{x}_{j}},\varphi_{\mathbf{x}_{j}},\theta_{\mathbf{y}_{j}},\varphi_{\mathbf{y}_{j}})
\end{equation}
\end{widetext}
where all quantities relative to the $j$ Faddeev component are functions of $\theta$ through eqs.~\eqref{eqs:jacobis}.

Local operators have a density vanishing in $\mathcal{O}(1/N)$, where $N = n_{c} n_{\alpha} n_{y} n_{x}$ is the total dimension of the operators, due to their banded structure. The density of the kernel operator is of the order of a few percent in our experience.

Finally, we work out explicitly the simplification arising from identical particles as well as the expression of the total wavefunction $\Psi$. When expressing the Schrödinger equation in terms of block operators for the different Faddeev components, we have
\begin{widetext}
\begin{align}
    \text{ for 3 types of particles} & \begin{cases}
        \left[
    \begin{pmatrix}
        \mathbb{t}_{1} + \mathbb{v}_{1} -E \mathbb{1}_{1} & \mathbb{0} & \mathbb{0} \\
        \mathbb{0} & \mathbb{t}_{2} + \mathbb{v}_{2} -E \mathbb{1}_{2} & \mathbb{0} \\
        \mathbb{0} & \mathbb{0} & \mathbb{t}_{3} + \mathbb{v}_{3} -E \mathbb{1}_{3}
    \end{pmatrix} + \begin{pmatrix}
        \mathbb{0} & \mathbb{K}_{12} & \mathbb{K}_{13} \\
        \mathbb{K}_{21} & \mathbb{0} & \mathbb{K}_{23} \\
        \mathbb{K}_{31} & \mathbb{K}_{32} & \mathbb{0} 
    \end{pmatrix}
    \right] \begin{pmatrix}
        \phi_{1} \\
        \phi_{2} \\
        \phi_{3}
    \end{pmatrix} = 0 \\
    \Psi =\phi_{1}(\mathbf{x}_{1},\mathbf{y}_{1}) + \phi_{2}(\mathbf{x}_{2},\mathbf{y}_{2}) + \phi_{3}(\mathbf{x}_{3},\mathbf{y}_{3})
    \end{cases} \\
        \text{ for 2 types of particles\footnotemark} & \begin{cases}
      \left[
     \begin{pmatrix}
         \mathbb{t}_{1} + \mathbb{v}_{1} -E \mathbb{1}_{1} & \mathbb{0} \\
         \mathbb{0} & \mathbb{t}_{2} + \mathbb{v}_{2} -E \mathbb{1}_{2} \\
     \end{pmatrix} + \begin{pmatrix}
         \mathbb{0} & 2 \mathbb{K}_{12} \\
         \mathbb{K}_{21} & \mathbb{K}_{23} \\
     \end{pmatrix}
     \right] \begin{pmatrix}
         \phi_{1} \\
         \phi_{2} \\
     \end{pmatrix} = 0 \\
    \Psi =\phi_{1}(\mathbf{x}_{1},\mathbf{y}_{1}) + \phi_{2}(\mathbf{x}_{2},\mathbf{y}_{2}) + \phi_{2}(\mathbf{x}_{3},\mathbf{y}_{3})
    \end{cases} \\
        \text{ for 1 type of particle} & \begin{cases}
      \left[
     \begin{pmatrix}
         \mathbb{t}_{1} + \mathbb{v}_{1} -E \mathbb{1}_{1}
     \end{pmatrix} + \begin{pmatrix}
         2\mathbb{K}_{12} 
     \end{pmatrix}
     \right] \begin{pmatrix}
         \phi_{1}
     \end{pmatrix} = 0 \\
    \Psi =\phi_{1}(\mathbf{x}_{1},\mathbf{y}_{1}) + \phi_{1}(\mathbf{x}_{2},\mathbf{y}_{2}) + \phi_{1}(\mathbf{x}_{3},\mathbf{y}_{3})
    \end{cases}
\end{align}
\addtocounter{footnote}{-1}
\footnotetext{We consider particles 2 and 3 identical.}
\end{widetext}

% section appendix_orthogonal_collocation_method (end)

% ***********Bibliography*********************

% \bibliographystyle{unsrt}
% \bibliography{TBR}

\end{document}